\documentclass[english,notitlepage,twocolumn,showpacs,pra]{revtex4-1}
\usepackage[T1]{fontenc}
\usepackage[latin9]{inputenc}
\usepackage{amsmath}
\usepackage{amssymb}
\usepackage{esint}
\usepackage{babel}
\begin{document}

\title{Large-momentum distribution of a polarized Fermi gas and $p$-wave
contacts}

\author{Shi-Guo Peng}
\email{pengshiguo@gmail.com}

\affiliation{State Key Laboratory of Magnetic Resonance and Atomic and Molecular
Physics, Wuhan Institute of Physics and Mathematics, Chinese Academy
of Sciences, Wuhan 430071, China}

\affiliation{Centre for Quantum and Optical Science, Swinburne University of Technology,
Melbourne 3122, Australia}

\author{Xia-Ji Liu}

\affiliation{Centre for Quantum and Optical Science, Swinburne University of Technology,
Melbourne 3122, Australia}

\author{Hui Hu}

\affiliation{Centre for Quantum and Optical Science, Swinburne University of Technology,
Melbourne 3122, Australia}

\date{\today}
\begin{abstract}
We present a derivation of the adiabatic energy relations as well
as the large momentum distribution of a polarized Fermi gas near $p$-wave
Feshbach resonances. The leading asymptotic behavior ($k^{-2}$) and
subleading behavior ($k^{-4}$) of the large momentum distribution
have recently been predicted by Yu \textit{et al.} {[}Phys. Rev. Lett.
\textbf{115}, 135304 (2015){]} and by He \textit{et al}. {[}Phys.
Rev. Lett. \textbf{116}, 045301 (2016){]} using two different approaches.
Here, we show that the subleading asymptotic behavior ($\sim k^{-4}$)
can not fully be captured by the contact defined from the adiabatic
energy relation related to the $p$-wave effective range, and there
should be an extra term resulted from the center-of-mass motion of
the pairs. The omission of this extra term is perhaps a reasonable
approximation at zero temperature. However, it should be taken into
account at finite temperature and should be of significant importance
to understand the recently measured momentum distribution in a resonant
$p$-wave Fermi gas of ultracold $^{40}$K atoms {[}Luciuk \textit{et
al.}, Nature Phys. \textbf{12}, 599 (2016){]}.
\end{abstract}

\maketitle

\section{Introduction}

In ultracold atomic physics, some key properties of a many-particle
system are governed by a set of universal relations that follow from
the short-range behavior of the two-body physics. For strongly interacting
two-component Fermi gases with $s$-wave contact interactions, in
his seminal works \cite{Tan2008E,Tan2008L,Tan2008G}, Tan discovered
the intrinsic connection between the $k^{-4}$ tail of the momentum
distribution at large $k$ and the derivative of the total energy
of the system with respect to the inverse $s$-wave scattering length,
both of which are related to a quantity named \emph{contact} \cite{Tan2008E,Tan2008L,Tan2008G}.
Afterwards, more universal Tan relations were derived within the concept
of contact \cite{Zwerger2011B}. Over the past decade, an impressive
amount of both theoretical and experimental efforts have been devoted
to confirm such universal Tan relations and explore their important
consequences \cite{Punk2007T,Braaten2008E,Zhang2009U,Schneider2010U,Werner2012G,Hu2010S,Stewart2010V,Kuhnle2010U,Sagi2012M,Hoinka2013P}.

For a polarized Fermi gas, in which all the atoms are in the same
hyperfine state, the low-energy collision is dominated by the $p$-wave
interaction channel due to its unique statistical properties, as experimentally
observed in $^{6}$Li \cite{Zhang2004P} as well as in $^{40}$K \cite{Ticknor2004M,Gunter2005P}.
The non-zero orbital angular momentum ($l=1$) of the $p$-wave interaction
leads to a spatial anisotropic scattering, and opens a new way to
manipulate resonant interatomic interactions by using a magnetic field
vector \cite{Peng2014M,Gao2015T}. Recently, the universal relations
near a $p$-wave resonance attract a great deal of attention \cite{Yu2015U,Yoshida2015U,Yoshida2016P,He2016C,Zhang2016C,Luciuk2016E}.
Unlike the $s$-wave case, there are two $p$-wave contacts involved,
which are related to the adiabatic variation of the energy with respect
to the scattering volume and to the effective range, and also capture
the leading and subleading asymptotic behavior of the large momentum
distribution, i.e, the $k^{-2}$ and $k^{-4}$ tails, respectively
\cite{Yu2015U}. These $p$-wave universal relations have recently
confirmed experimentally, although the measured contact related to
the effective range can not be explained theoretically \cite{Luciuk2016E}.
Furthermore, the $p$-wave contacts are also defined based on a two-channel
model \cite{Yoshida2015U,Yoshida2016P}. Besides, He \emph{et al}.
defined the contacts according to the large-momentum distribution,
based on the removing of the divergence of the internal energy, then
more contacts would appear if higher partial waves are included \cite{He2016C,Zhang2016C,Zhang2016E}.
Currently, the research on the $p$-wave universal relations grows
rapidly, and all the above progress is made within less than one year.
Among these universal relations, the adiabatic energy theorem and
large momentum distribution are of particular interest, since they
manifest in an elegant way how two seemingly uncorrelated observables
- the thermodynamics and the short-range behavior of a many-particle
system - are related. 

In this work, we present a \emph{rigorous} derivation of the adiabatic
energy theorem as well as the large momentum distribution of a polarized
Fermi gas with $p$-wave interaction, following the route of Tan's
original work about the $s$-wave case \cite{Tan2008E,Tan2008L}.
We define two $p$-wave contacts as in the previous work \cite{Yu2015U}
by using the adiabatic energy theorems, i.e., $\mathcal{C}_{a}$ and
$\mathcal{C}_{b}$, which are proportional to the derivative of the
total energy with respect to the inverse $p$-wave scattering volume
and to the effective range, respectively. We show that the contact
$\mathcal{C}_{a}$ related to the scattering volume completely captures
the leading asymptotic behavior of the large momentum distribution
$\mathcal{C}_{a}/k^{2}$, a result that fully agrees with those of
\cite{Yu2015U,Yoshida2015U}. However, in the subleading behavior
at the order of $k^{-4}$, we find that there is an \emph{extra} term,
in addition to the contact $\mathcal{C}_{b}$ term related to the
effective range. This additional term is resulted from the center-of-mass
(c.m.) motion of the pairs, and was unfortunately missed in the previous
work \cite{Yu2015U}. Such an extra term should be taken into account
at finite temperature, and might be non-negligible even at zero temperature,
due to the large pair fluctuations near the resonance. It gives the
hint why the experimentally measured $\mathcal{C}_{b}$ can not be
understood by using the same theoretical model, which otherwise provides
a good qualitative explanation for the measured $\mathcal{C}_{a}$
\cite{Luciuk2016E}.

This paper is arranged as follows. To warm up, in the next section
(Sec. \ref{sec:SWaveContact}), we briefly review the definitions
of the $s$-wave contacts, and introduce an additional contact due
to the $s$-wave effective range. We derive the corresponding adiabatic
energy theorems, and show their connection to the tails of the large
momentum distribution. The non-trivial consequence of the effective
range is highlighted. In Sec. \ref{sec:AdiabaticEnergyTheorem}, we
derive two adiabatic energy theorems of a polarized Fermi gas with
$p$-wave interaction, and define two contacts associated with the
scattering volume and effective range. The asymptotic behavior of
the large momentum distribution near $p$-wave resonances is discussed
in Sec. \ref{sec:LargeMomentumDistribution}. We show how the extra
term resulted from the c.m. motion of the pairs appears in the subleading
order ($k^{-4}$), in addition to the contact term related to the
effective range. Finally, our main results are summarized in Sec.
\ref{sec:Conclusions}.

\section{Review on the $s$-wave contacts\label{sec:SWaveContact}}

Before the discussion of the $p$-wave contacts, we review the definition
of the $s$-wave contact, and consider a more general case, in which
the finite-range effect is included. This case has been previously
addressed by Werner and Castin \cite{Werner2012G}. In addition to
confirming and generalizing their results, the derivation presented
below clarify some technical details, which turn out to be crucial
in handling the more complicated $p$-wave case. 

Let us consider a strongly interacting two-component atomic Fermi
gas confined in an external potential, with $N_{\uparrow}$ spin-up
and $N_{\downarrow}$ spin-down fermions, respectively. The interatomic
interaction between two fermions with unlike spins is assumed to be
short range with a characteristic length $\epsilon$, which is much
smaller than the length scale of the external potential $\eta$. Then
we may deal with the interaction by setting a short-range boundary
condition on the many-body wavefunction: For any pair of fermions
$i$ and $j$ in different spin states, there exists a \emph{regular}
function $\mathcal{A}\left(\mathbf{X},\mathbf{R}\right)$ \cite{Werner2012G},
and the many-body wavefunction can be decoupled as 
\begin{equation}
\Psi\left(\mathbf{X},\mathbf{R},\mathbf{r}\right)\approx\mathcal{A}\left(\mathbf{X},\mathbf{R}\right)\psi\left(\mathbf{r}\right),\label{eq:2.1}
\end{equation}
when the fermions $i,\,j$ get close enough to each other but still
far away compared to the range of the interaction potential, i.e.,
$\epsilon\ll r\ll\eta$. Here, $\mathbf{r}=\mathbf{r}_{i}-\mathbf{r}_{j}$
and $\mathbf{R}=\left(\mathbf{r}_{i}+\mathbf{r}_{j}\right)/2$ are
respectively the relative and c.m. coordinates of the pair $\left(i,j\right)$,
$\mathbf{X}$ includes those of the rest of the fermions, and $\psi\left(\mathbf{r}\right)$
can approximately be treated as the relative wavefunction of the pair.
The regularity of the function $\mathcal{A}\left(\mathbf{X},\mathbf{R}\right)$
means that no more pairs except the fermions $i$ and $j$ can interact
with each other, according to such a short-range interaction. 

\subsection{Adiabatic energy theorems}

For two many-body wavefunction $\Psi$ and $\Psi^{\prime}$ corresponding
to different interatomic interaction strengths, they satisfy the Schr\"{o}dinger
equation with different energies
\begin{eqnarray}
\sum_{n=1}^{N}\left[-\frac{\hbar^{2}}{2M}\nabla_{n}^{2}+U\left(\mathbf{r}_{n}\right)\right]\Psi & = & E\Psi,\label{eq:2.2}\\
\sum_{n=1}^{N}\left[-\frac{\hbar^{2}}{2M}\nabla_{n}^{2}+U\left(\mathbf{r}_{n}\right)\right]\Psi^{\prime} & = & E^{\prime}\Psi^{\prime},\label{eq:2.3}
\end{eqnarray}
if there is no pair with different spin states within the range of
the interaction. Here, $N$ is the total atom number, $M$ is the
atomic mass, and $U\left(\mathbf{r}_{n}\right)$ is the external potential
experienced by the $n$-th fermion. By subtracting $\Psi\times$ {[}Eq.
(\ref{eq:2.3}){]}$^{*}$ from $\Psi^{\prime*}\times$ Eq. (\ref{eq:2.2}),
and integrating over the domain $\mathcal{D}_{\epsilon}$, the set
of all configurations $\left(\mathbf{r}_{i},\mathbf{r}_{j}\right)$
in which $r=\left|\mathbf{r}_{i}-\mathbf{r}_{j}\right|>\epsilon$
\cite{Tan2008L,Werner2012G}, we arrive at
\begin{multline}
\left(E-E^{\prime}\right)\int_{\mathcal{D}_{\epsilon}}\Psi^{\prime*}\Psi d\mathbf{X}d\mathbf{R}d\mathbf{r}=\\
-\frac{\hbar^{2}}{M}\mathcal{N}\oiint_{r=\epsilon}\left(\Psi^{\prime*}\nabla_{\mathbf{r}}\Psi-\Psi\nabla_{\mathbf{r}}\Psi^{\prime*}\right)\cdot\hat{\mathbf{n}}d\mathcal{S},\label{eq:2.4}
\end{multline}
where $\mathcal{N}\equiv N_{\uparrow}N_{\downarrow}$ is the number
of all the possible ways to pair atoms, $\mathcal{S}$ is the surface
that the distance between the two atoms in the pair $\left(i,j\right)$
is $\epsilon$ , and $\hat{\mathbf{n}}$ is the direction normal to
$\mathcal{S}$, but is opposite to the radial direction.

For the $s$-wave interaction, when $\epsilon\ll r\ll\eta$, the many-body
wavefunction $\Psi\left(\mathbf{X},\mathbf{R},\mathbf{r}\right)$
in Eq. (\ref{eq:2.1}) takes the form of
\begin{equation}
\Psi\left(\mathbf{X},\mathbf{R},\mathbf{r}\right)\approx\mathcal{A}\left(\mathbf{X},\mathbf{R}\right)\cdot k\left[j_{0}\left(kr\right)\cot\delta_{0}-n_{0}\left(kr\right)\right],\label{eq:2.5}
\end{equation}
where $j_{l}\left(\cdot\right),\,n_{l}\left(\cdot\right)$ are the
spherical Bessel functions of the first and the second kinds, $\delta_{0}$
is the $s$-wave scattering phase shift, $k$ is the relative wavenumber
of the pair $\left(i,j\right)$ , i.e, $\hbar^{2}k^{2}/M=\varepsilon\left(\mathbf{X},\mathbf{R}\right)$
\cite{Werner2012G}, 
\begin{equation}
\varepsilon\left(\mathbf{X},\mathbf{R}\right)\equiv E-\frac{1}{\mathcal{A}\left(\mathbf{X},\mathbf{R}\right)}\left[T\left(\mathbf{X},\mathbf{R}\right)+U\left(\mathbf{X},\mathbf{R}\right)\right]\mathcal{A}\left(\mathbf{X},\mathbf{R}\right),\label{eq:2.6}
\end{equation}
and $T\left(\mathbf{X},\mathbf{R}\right)$ and $U\left(\mathbf{X},\mathbf{R}\right)$
are respectively the kinetic and external potential operators including
the c.m. motion of the pair $\left(i,j\right)$ and those of the rest
of the fermions. Expanding at $r\sim0^{+}\left(\gg\epsilon\right)$,
we easily obtain the following asymptotic form of the many-body wavefunction
\begin{equation}
\Psi\left(\mathbf{X},\mathbf{R},\mathbf{r}\right)=\mathcal{A}\left(\mathbf{X},\mathbf{R}\right)\left(\frac{1}{r}-\frac{1}{a}+\frac{1}{2}bk^{2}\right)+\mathcal{O}\left(r\right),\label{eq:2.7}
\end{equation}
where we have used the effective-range expansion of the $s$-wave
scattering phase shift at small $k$,
\begin{equation}
k\cot\delta_{0}=-\frac{1}{a}+\frac{1}{2}bk^{2}+\mathcal{O}\left(k^{4}\right),\label{eq:2.8}
\end{equation}
where $a,\,b$ are the $s$-wave scattering length and effective range,
respectively. Inserting the asymptotic form of the many-body wavefunction
(\ref{eq:2.7}) into Eq. (\ref{eq:2.4}), we find
\begin{multline}
\left(E-E^{\prime}\right)\int_{\mathcal{D}_{\epsilon}}\Psi^{\prime*}\Psi d\mathbf{X}d\mathbf{R}d\mathbf{r}\\
=-\frac{4\pi\hbar^{2}}{M}I_{a}\left(\frac{1}{a}-\frac{1}{a^{\prime}}\right)+2\pi\left(\bar{\varepsilon}b-\bar{\varepsilon}^{\prime}b^{\prime}\right),\label{eq:2.9}
\end{multline}
where 
\begin{eqnarray}
\bar{\varepsilon} & = & \mathcal{N}\int d\mathbf{X}d\mathbf{R}\mathcal{A}^{\prime*}\left[E-T-U\right]\mathcal{A},\label{eq:2.10}\\
\bar{\varepsilon}^{\prime} & = & \mathcal{N}\int d\mathbf{X}d\mathbf{R}\mathcal{A}\left[E^{\prime}-T-U\right]\mathcal{A}^{\prime*},\label{eq:2.11}
\end{eqnarray}
and
\begin{equation}
I_{a}\equiv\mathcal{N}\int d\mathbf{X}d\mathbf{R}\mathcal{A}^{\prime*}\left(\mathbf{X},\mathbf{R}\right)\mathcal{A}\left(\mathbf{X},\mathbf{R}\right).\label{eq:2.12}
\end{equation}
Let $E^{\prime}\rightarrow E$ , $a^{\prime}\rightarrow a$ , and
$b^{\prime}\rightarrow b$, then we find
\begin{multline}
\delta E\cdot\int_{\mathcal{D}_{\epsilon}}\left|\Psi\right|^{2}d\mathbf{X}d\mathbf{R}d\mathbf{r}\\
=-\frac{4\pi\hbar^{2}}{M}\mathcal{I}_{a}\cdot\delta a^{-1}+2\pi\bar{\varepsilon}\cdot\delta b+2\pi\mathcal{I}_{a}b\cdot\delta E,\label{eq:2.13}
\end{multline}
and 
\begin{equation}
\mathcal{I}_{a}\equiv\mathcal{N}\int d\mathbf{X}d\mathbf{R}\left|\mathcal{A}\left(\mathbf{X},\mathbf{R}\right)\right|^{2}.\label{eq:2.14}
\end{equation}
If the finite-range effect cannot be neglected, i.e, $b\neq0$ but
small, we find (see Appendix A)
\begin{equation}
\int_{\mathcal{D}_{\epsilon}}\left|\Psi\right|^{2}d\mathbf{X}d\mathbf{R}d\mathbf{r}\approx1+2\pi\mathcal{I}_{a}b,\label{eq:2.15}
\end{equation}
and then
\begin{equation}
\delta E=-\frac{4\pi\hbar^{2}}{M}\mathcal{I}_{a}\cdot\delta a^{-1}+2\pi\bar{\varepsilon}\cdot\delta b,\label{eq:2.16}
\end{equation}
which yields
\begin{eqnarray}
\frac{\partial E}{\partial a^{-1}} & = & -\frac{4\pi\hbar^{2}}{M}\mathcal{I}_{a},\label{eq:2.17}\\
\frac{\partial E}{\partial b} & = & 2\pi\bar{\varepsilon},\label{eq:2.18}
\end{eqnarray}
and $\bar{\varepsilon}$ becomes
\begin{equation}
\bar{\varepsilon}=\mathcal{N}\int d\mathbf{X}d\mathbf{R}\mathcal{A}^{*}\left[E-T-U\right]\mathcal{A}.\label{eq:2.19-1}
\end{equation}
We note that, the normalization of the wave-function Eq. (\ref{eq:2.15})
- derived by us in Appendix A - is crucial to obtain the adiabatic
energy relation. In the previous work \cite{Werner2012G}, Werner
and Castin considered the finite-range correction in the limit of
$b=0$ only, where the third term on the right-hand-side of Eq. (\ref{eq:2.13})
drops out automatically. 

\subsection{Large momentum distribution}

Next, let us consider the asymptotic behavior of the large momentum
distribution of the system. The momentum distribution of the system
for spin-up fermions is defined as
\begin{eqnarray}
n_{\uparrow}\left(\mathbf{k}\right) & \equiv & \sum_{i=1}^{N_{\uparrow}}\int\prod_{t\neq i}d\mathbf{r}_{t}\left|\int d\mathbf{r}_{i}\Psi e^{-i\mathbf{k}\cdot\mathbf{r}_{i}}\right|^{2}.\label{eq:2.19}
\end{eqnarray}
When the $i$th spin-up and the $j$th spin-down fermions are close,
we formally expand the many-body wavefunction at the short range $r\sim0^{+}\left(\gg\epsilon\right)$
up to the order $\mathcal{O}(r)$ as
\begin{multline}
\Psi\left(\mathbf{X},\mathbf{R},\mathbf{r}\right)=\frac{\mathcal{A}\left(\mathbf{X},\mathbf{R}\right)}{r}+\mathcal{B}\left(\mathbf{X},\mathbf{R}\right)r\\
+\mathcal{C}\left(\mathbf{X},\mathbf{R}\right)+\mathbf{r}\cdot\mathbf{L}\left(\mathbf{X},\mathbf{R}\right)+\mathcal{O}\left(r^{2}\right),\label{eq:2.20}
\end{multline}
where $\mathcal{A},\,\mathcal{B},\,\mathcal{C}$ and $\mathbf{L}$
are all regular functions, and the last term in the above, i.e., $\mathbf{r}\cdot\mathbf{L}\left(\mathbf{X},\mathbf{R}\right)$,
represents the coupling between the relative and c.m. motions of the
pair $\left(i,j\right)$, resulted from the external confinement.
Comparing Eqs. (\ref{eq:2.5}) and (\ref{eq:2.20}) at small $r$,
we easily find
\begin{eqnarray}
\mathcal{B}\left(\mathbf{X},\mathbf{R}\right) & = & -\mathcal{A}\left(\mathbf{X},\mathbf{R}\right)\cdot\frac{M\varepsilon\left(\mathbf{X},\mathbf{R}\right)}{2\hbar^{2}},\label{eq:2.21}\\
\mathcal{C}\left(\mathbf{X},\mathbf{R}\right) & = & \mathcal{A}\left(\mathbf{X},\mathbf{R}\right)\left[-\frac{1}{a}+b\cdot\frac{M\varepsilon\left(\mathbf{X},\mathbf{R}\right)}{2\hbar^{2}}\right],\label{eq:2.22}
\end{eqnarray}
or the constraint on the expansion coefficients $\mathcal{A},\,\mathcal{B},\,\mathcal{C}$:
\begin{equation}
\frac{\mathcal{A}\left(\mathbf{X},\mathbf{R}\right)}{a}+b\mathcal{B}\left(\mathbf{X},\mathbf{R}\right)+\mathcal{C}\left(\mathbf{X},\mathbf{R}\right)=0,\label{eq:2.23}
\end{equation}
which is an alternative expression of the short-range boundary condition
for the many-body wavefunction. In addition, $\bar{\varepsilon}$
in Eq.(\ref{eq:2.19-1}) can be written as
\begin{equation}
\bar{\varepsilon}=-\frac{2\hbar^{2}}{M}\mathcal{N}\int d\mathbf{X}d\mathbf{R}\mathcal{A}^{*}\left(\mathbf{X},\mathbf{R}\right)\mathcal{B}\left(\mathbf{X},\mathbf{R}\right)\equiv-\frac{2\hbar^{2}}{M}\mathcal{I}_{b},\label{eq:2.24}
\end{equation}
where
\begin{equation}
\mathcal{I}_{b}\equiv\mathcal{N}\int d\mathbf{X}d\mathbf{R}\mathcal{A}^{*}\left(\mathbf{X},\mathbf{R}\right)\mathcal{B}\left(\mathbf{X},\mathbf{R}\right)\label{eq:2.25}
\end{equation}
should be real, apparently. In the definition of the momentum distribution
(\ref{eq:2.19}), we may rewrite $n_{\uparrow}\left(\mathbf{k}\right)$
as
\begin{equation}
n_{\uparrow}\left(\mathbf{k}\right)=\sum_{i=1}^{N_{\uparrow}}\int\prod_{t\neq i}d\mathbf{r}_{t}\left|\tilde{\Psi}_{i}\left(\mathbf{k}\right)\right|^{2},\label{eq:2.26}
\end{equation}
where $\tilde{\Psi}_{i}\left({\bf k}\right)\equiv\int d\mathbf{r}_{i}\Psi e^{-i\mathbf{k}\cdot\mathbf{r}_{i}}$
. In the large-$k$ limit, we know that the Fourier transform with
respect to ${\bf r}_{i}$ is dominated by the behavior of the wavefunction
at short distances between the atom $i$ and the other atoms, then
we have \cite{Werner2012G} 
\begin{equation}
\tilde{\Psi}_{i}\left(\mathbf{k}\right)\underset{k\rightarrow\infty}{\approx}\sum_{j=1}^{N_{\downarrow}}e^{-i\mathbf{k}\cdot\mathbf{r}_{j}}\int d\mathbf{r}\Psi\left(\mathbf{X},\mathbf{r}_{j}+\frac{\mathbf{r}}{2},\mathbf{r}\right)e^{-i\mathbf{k}\cdot\mathbf{r}}.\label{eq:2.27}
\end{equation}
 Using $\nabla^{2}\left(r^{-1}\right)=-4\pi\delta\left(\mathbf{r}\right)$,
we have the identity
\begin{equation}
f\left(k\right)\equiv\int d\mathbf{r}\frac{e^{-i\mathbf{k}\cdot\mathbf{r}}}{r}=\frac{4\pi}{k^{2}},\label{eq:2.28}
\end{equation}
 so that 
\begin{eqnarray}
 &  & \int d\mathbf{r}\frac{\mathcal{A}\left(\mathbf{X},\mathbf{r}_{j}+{\bf r}/2\right)}{r}e^{-i\mathbf{k}\cdot\mathbf{r}}\nonumber \\
 & \approx & \mathcal{A}\left(\mathbf{X},\mathbf{r}_{j}\right)f\left(k\right)+i\frac{\nabla_{\mathbf{r}_{j}}\mathcal{A}\left(\mathbf{X},\mathbf{r}_{j}\right)}{2}\cdot\nabla_{{\bf k}}f\left(k\right)\nonumber \\
 &  & -\frac{1}{8}\left(\nabla_{{\bf r}_{j}}\cdot\nabla_{{\bf k}}\right)\left[\nabla_{{\bf r}_{j}}\mathcal{A}\left(\mathbf{X},\mathbf{r}_{j}\right)\cdot\nabla_{{\bf k}}f\left(k\right)\right]\nonumber \\
 & = & \mathcal{A}\left(\mathbf{X},\mathbf{r}_{j}\right)\frac{4\pi}{k^{2}}-i\left(\nabla_{{\bf r}_{j}}\mathcal{A}\cdot\hat{{\bf k}}\right)\frac{4\pi}{k^{3}}\nonumber \\
 &  & +\left[k^{4}\left(\nabla_{{\bf r}_{j}}\cdot\nabla_{{\bf k}}\right)\frac{\nabla_{{\bf r}_{j}}\mathcal{A}\cdot\hat{{\bf k}}}{k^{3}}\right]\frac{\pi}{k^{4}}+\mathcal{O}\left(k^{-5}\right),\label{eq:2.29}
\end{eqnarray}
 where $\hat{{\bf k}}$ is the unit vector of the radial direction
of ${\bf k}$ , and
\begin{equation}
\int d\mathbf{r}\mathcal{B}\left(\mathbf{X},{\bf r}_{j}+{\bf r}/2\right)re^{-i\mathbf{k}\cdot\mathbf{r}}\approx-\mathcal{B}\left(\mathbf{X},\mathbf{r}_{j}\right)\frac{8\pi}{k^{4}}+\mathcal{O}\left(k^{-5}\right).\label{eq:2.30}
\end{equation}
We can easily verify that the terms $\mathcal{C}(\mathbf{X},{\bf r}_{j}+{\bf r}/2)$
and $\mathbf{r}\cdot\mathbf{L}(\mathbf{X},{\bf r}_{j}+{\bf r}/2)$
contribute only to the small momentum other than the large-momentum
tail. By substituting Eqs. (\ref{eq:2.29}) and (\ref{eq:2.30}) into
Eq. (\ref{eq:2.27}), and then into Eq. (\ref{eq:2.26}), we finally
obtain the asymptotic behavior of $n_{\uparrow}\left(\mathbf{k}\right)$
at $k\rightarrow\infty$,
\begin{multline}
n_{\uparrow}\left(\mathbf{k}\right)\sim\frac{\mathcal{C}_{a}}{k^{4}}+\text{{\bf Im}}\left[\mathcal{N}\int d\mathbf{X}d\mathbf{R}\mathcal{A}^{*}\left(\nabla_{{\bf R}}\mathcal{A}\cdot\hat{{\bf k}}\right)\right]\frac{32\pi^{2}}{k^{5}}\\
+\left\{ \mathcal{C}_{b}+8\pi^{2}\mathcal{N}\int d{\bf X}d{\bf R}\left[6\left|\nabla_{{\bf R}}\mathcal{A}\cdot\hat{{\bf k}}\right|^{2}-\left|\nabla_{{\bf R}}\mathcal{A}\right|^{2}\right]\right\} \frac{1}{k^{6}}\\
+\mathcal{O}\left(k^{-7}\right),\label{eq:2.31}
\end{multline}
 where we have defined two contacts as
\begin{eqnarray}
\mathcal{C}_{a} & \equiv & 16\pi^{2}\mathcal{I}_{a},\label{eq:2.32}\\
\mathcal{C}_{b} & \equiv & -64\pi^{2}\mathcal{I}_{b},\label{eq:2.33}
\end{eqnarray}
and have set $\mathbf{r}_{j}=\mathbf{R}$. If we take the average
of the momentum distribution over the direction of ${\bf k}$ , i.e.,
$\int d\hat{{\bf k}}n\left({\bf k}\right)/4\pi$ , we find the odd-order
terms of $k^{-1}$ vanish (for example, the coefficient of $k^{-5}$
term is simply proportional to $\cos\theta_{{\bf k}}$ , where $\theta_{{\bf k}}$
is the polar angle of ${\bf k}$, and then vanishes after the integration
over $\theta_{{\bf k}}$). In the $k^{-6}$ term, since $\int d\hat{{\bf k}}\cos^{2}\theta_{{\bf k}}/4\pi=1/3$
, we find $\left|\nabla_{{\bf R}}\mathcal{A}\cdot\hat{{\bf k}}\right|^{2}$
becomes $\left|\nabla_{{\bf R}}\mathcal{A}\right|^{2}/3$ after taking
the average. Then the average of the momentum distribution becomes
\begin{multline}
n_{\uparrow}\left(k\right)\sim\frac{\mathcal{C}_{a}}{k^{4}}+\left(\mathcal{C}_{b}+8\pi^{2}\mathcal{N}\int d{\bf X}d{\bf R}\left|\nabla_{{\bf R}}\mathcal{A}\right|^{2}\right)\cdot\frac{1}{k^{6}}\\
+\mathcal{O}\left(k^{-8}\right),\label{eq:2.34}
\end{multline}
which exactly agrees with the result of \cite{Werner2012G}. Obviously,
there is an extra term besides the contact $\mathcal{C}_{b}$ appearing
in the coefficient of the subleading behavior $1/k^{-6}$, which is
resulted from the c.m. motion of the pairs \cite{Werner2012G}. Physically,
for a pair of fermions $\uparrow$ and $\downarrow$, the wavevector
$\mathbf{k}_{\uparrow}$ is the linear combination of the relative
wavevector $\mathbf{k}$ and the c.m. wavevector $\mathbf{K}$, i.e.,
$\mathbf{k}_{\uparrow}=\mathbf{k}+\mathbf{K}/2$. Therefore, a nonzero
$\mathbf{K}$ provides an extra $k_{\uparrow}^{-6}$ subleading contribution
to the single-particle momentum distribution $n_{\uparrow}\left(k_{\uparrow}\right)$,
even if the probability distribution $n_{\uparrow}\left(k\right)$
is exactly scaled as $k^{-4}$. 

Using the definitions of the contacts, i.e, Eqs. (\ref{eq:2.32})
and (\ref{eq:2.33}), the adiabatic energy theorems (\ref{eq:2.17})
and (\ref{eq:2.18}) can be rewritten as
\begin{eqnarray}
\frac{\partial E}{\partial a^{-1}} & = & -\frac{\hbar^{2}\mathcal{C}_{a}}{4\pi M},\label{eq:2.35}\\
\frac{\partial E}{\partial b} & = & \frac{\hbar^{2}\mathcal{C}_{b}}{16\pi M}.\label{eq:2.36}
\end{eqnarray}
Eq.(\ref{eq:2.35}) is the well-known $s$-wave adiabatic energy theorem
derived by Tan \cite{Tan2008L} at the zero-range limit. However,
if the finite-range effect is included, there is an additional adiabatic
energy theorem (\ref{eq:2.36}) related to the effective range \cite{Werner2012G},
and the contact $\mathcal{C}_{b}$ appears.

However, it is obvious from Eq. (\ref{eq:2.34}) that in the presence
of a finite effective range, the subleading tail ($k^{-6}$) can not
be simply described by the contact $\mathcal{C}_{b}$. As we shall
see below, this turns out be a very general feature and happens to
the $p$-wave interaction as well.

\section{$p$-wave adiabatic energy theorems\label{sec:AdiabaticEnergyTheorem}}

The derivation of the $s$-wave universal relations in the above can
be directly generalized to spin-polarized Fermi gases with $p$-wave
interatomic interactions. When the distance between the fermions $i,j$
becomes small, while the other fermions are all far away from each
other, the many-body wavefunction takes the form \cite{Weiss2015G}
\begin{multline}
\Psi\left(\mathbf{X},\mathbf{R},\mathbf{r}\right)\approx\mathcal{A}\left(\mathbf{X},\mathbf{R}\right)\times\\
\sum_{m=-1}^{1}G_{m}k^{2}\left[j_{1}\left(kr\right)\cot\delta_{1m}-n_{1}\left(kr\right)\right]Y_{1m}\left(\hat{\mathbf{r}}\right),\label{eq:3.1}
\end{multline}
where $\delta_{1m}$ is the $p$-wave scattering phase shift corresponding
to the magnetic quantum number $m$, $Y_{lm}\left(\hat{\mathbf{r}}\right)$
is the spherical harmonics, $k$ is the relative wavenumber of the
pair $(i,j)$ defined similarly to that of the $s$-wave case, and
$G_{m}$ is the expansion coefficient of the relative wavefunction
of the pair $\left(i,j\right)$. Expanding the many-body wavefunction
(\ref{eq:3.1}) at small $r$, we obtain 
\begin{multline}
\Psi\left(\mathbf{X},\mathbf{R},{\bf r}\right)\approx\mathcal{A}\left(\mathbf{X},\mathbf{R}\right)\sum_{m}G_{m}\left[\frac{1}{r^{2}}+\frac{k^{2}}{2}\right.\\
\left.+\left(\frac{b_{m}k^{2}}{6}-\frac{1}{3a_{m}}\right)r+\mathcal{O}\left(r^{2}\right)\right]Y_{1m}\left(\hat{\mathbf{r}}\right),\label{eq:3.2}
\end{multline}
where we have used the effective-range expansion of the $p$-wave
scattering phase shift,
\begin{equation}
k^{3}\cot\delta_{1m}=-\frac{1}{a_{m}}+\frac{1}{2}b_{m}k^{2}.\label{eq:3.3}
\end{equation}
We should note that the quantities $a_{m}$ and $b_{m}$ have the
dimensions of length$^{3}$ and length$^{-1}$. Therefore, $a_{m}$
is usually called the $p$-wave scattering volume, while $b_{m}$
is still called the effective range as a matter of convention. Here,
we assume the scattering volume $a_{m}$ and the effective range $b_{m}$
are dependent on $m$, which is true for $p$-wave collisions, such
as in $^{40}$K \cite{Ticknor2004M,Gunter2005P}. Inserting Eq. (\ref{eq:3.2})
into Eq. (\ref{eq:2.4}), we obtain
\begin{multline}
\left(E-E^{\prime}\right)\int_{\mathcal{D}_{\epsilon}}\Psi^{\prime*}\Psi d\mathbf{X}d\mathbf{R}d\mathbf{r}=\\
\sum_{m}G_{m}^{\prime*}G_{m}\left\{ -\frac{\hbar^{2}}{M}I_{a}\left(\frac{1}{a_{m}}-\frac{1}{a_{m}^{\prime}}\right)+\right.\\
\left.\left[\left(\frac{b_{m}}{2}+\frac{1}{\epsilon}\right)\bar{\varepsilon}-\left(\frac{b_{m}^{\prime}}{2}+\frac{1}{\epsilon}\right)\bar{\varepsilon}^{\prime}\right]\right\} ,\label{eq:3.4}
\end{multline}
where $\bar{\varepsilon}$, $\bar{\varepsilon}^{\prime}$, and $I_{a}$
are defined in the similar way as those in the $s$-wave case, i.e.,
Eqs. (\ref{eq:2.10})-(\ref{eq:2.12}), but with $\mathcal{N}=N\left(N-1\right)/2$
, and $N$ is the total number of atoms. Letting $E^{\prime}\rightarrow E$,
$a_{m}^{\prime}\rightarrow a_{m}$, and $b_{m}^{\prime}\rightarrow b_{m}$,
we obtain 
\begin{multline}
\delta E\int_{\mathcal{D}_{\epsilon}}\left|\Psi\right|^{2}d\mathbf{X}d\mathbf{R}d\mathbf{r}=\sum_{m}\left|G_{m}\right|^{2}\left\{ -\frac{\hbar^{2}}{M}\mathcal{I}_{a}\delta a_{m}^{-1}+\right.\\
\left.\frac{\bar{\varepsilon}}{2}\delta b_{m}+\mathcal{I}_{a}\left(\frac{b_{m}}{2}+\frac{1}{\epsilon}\right)\delta E\right\} ,\label{eq:3.5}
\end{multline}
where $\mathcal{I}_{a}$ is similarly defined (see Eq. (\ref{eq:2.14})).
This expression can be simplified by using the normalization of the
wavefunction,
\begin{equation}
\int_{\mathcal{D}_{\epsilon}}\left|\Psi\right|^{2}d\mathbf{X}d\mathbf{R}d\mathbf{r}\approx1+\mathcal{I}_{a}\sum_{m}\left|G_{m}\right|^{2}\left(\frac{b_{m}}{2}+\frac{1}{\epsilon}\right),\label{eq:3.6}
\end{equation}
which is discussed in Appendix A. We find
\begin{equation}
\delta E=\sum_{m}\left|G_{m}\right|^{2}\left(-\frac{\hbar^{2}}{M}\mathcal{I}_{a}\cdot\delta a_{m}^{-1}+\frac{\bar{\varepsilon}}{2}\cdot\delta b_{m}\right),\label{eq:3.7}
\end{equation}
and hence
\begin{eqnarray}
\frac{\partial E}{\partial a_{m}^{-1}} & = & -\frac{\hbar^{2}}{M}\mathcal{I}_{a}\left|G_{m}\right|^{2},\label{eq:3.8}\\
\frac{\partial E}{\partial b_{m}} & = & \frac{\bar{\varepsilon}}{2}\left|G_{m}\right|^{2}.\label{eq:3.9}
\end{eqnarray}

\section{Tail of the large momentum distribution near $p$-wave resonances\label{sec:LargeMomentumDistribution}}

The momentum distribution at large $k$ is determined by the short-range
behavior of the many-body wavefunction when the fermions $i$ and
$j$ are close. Similar to the $s$-wave case, we formally write the
many-body wavefunction as
\begin{multline}
\Psi\left(\mathbf{X},\mathbf{R},{\bf r}\right)\approx\sum_{m}G_{m}\left(\frac{\mathcal{A}}{r^{2}}+\mathcal{B}+\mathcal{C}r\right)Y_{1m}\left(\hat{\mathbf{r}}\right)\\
+\mathbf{r}\cdot\mathbf{L}+\mathcal{O}\left(r^{2}\right),\label{eq:4.1}
\end{multline}
where we have omitted the arguments $\mathbf{X},\mathbf{R}$ of the
functions $\mathcal{A},\,\mathcal{B},\,\mathcal{C}$ and $\mathbf{L}$
to simplify the expression. Expanding Eq. (\ref{eq:3.1}) at small
$r$, and comparing with Eq. (\ref{eq:4.1}), we easily obtain
\begin{eqnarray}
\mathcal{B}\left(\mathbf{X},\mathbf{R}\right) & = & \frac{M\varepsilon}{2\hbar^{2}}\mathcal{A}\left(\mathbf{X},\mathbf{R}\right),\label{eq:4.2}\\
\mathcal{C}\left(\mathbf{X},\mathbf{R}\right) & = & \left(\frac{b_{m}M\varepsilon}{6\hbar^{2}}-\frac{1}{3a_{m}}\right)\mathcal{A}\left(\mathbf{X},\mathbf{R}\right),\label{eq:4.3}
\end{eqnarray}
 or the constraint on the expansion coefficients $\mathcal{A},\,\mathcal{B},\,\mathcal{C}$:
\begin{equation}
\frac{\mathcal{A}\left(\mathbf{X},\mathbf{R}\right)}{3a_{m}}-\frac{b_{m}}{3}\mathcal{B}\left(\mathbf{X},\mathbf{R}\right)+\mathcal{C}\left(\mathbf{X},\mathbf{R}\right)=0,\label{eq:4.4}
\end{equation}
 which is an alternative expression of the short-range boundary condition
for the $p$-wave interaction \cite{Peng2014M}. Here, $\varepsilon$
is defined as in Eq. (\ref{eq:2.6}). Then we find
\begin{equation}
\bar{\varepsilon}=\frac{2\hbar^{2}}{M}\mathcal{N}\int d\mathbf{X}d\mathbf{R}\mathcal{A}^{*}\left(\mathbf{X},\mathbf{R}\right)\mathcal{B}\left(\mathbf{X},\mathbf{R}\right)\equiv\frac{2\hbar^{2}}{M}\mathcal{I}_{b},\label{eq:4.5}
\end{equation}
 where 
\begin{equation}
\mathcal{I}_{b}\equiv\mathcal{N}\int d\mathbf{X}d\mathbf{R}\mathcal{A}^{*}\left(\mathbf{X},\mathbf{R}\right)\mathcal{B}\left(\mathbf{X},\mathbf{R}\right).\label{eq:4.6}
\end{equation}

In the following, we derive the momentum distribution $n\left(\mathbf{k}\right)$
at large $k$ using its definition Eq. (\ref{eq:2.26}) and the many-body
wavefunction Eq. (\ref{eq:4.1}). With the help of the identities,
\begin{equation}
e^{i\mathbf{k}\cdot\mathbf{r}}=4\pi\sum_{lm}i^{l}j_{l}\left(kr\right)Y_{lm}^{*}\left(\hat{\mathbf{k}}\right)Y_{lm}\left(\hat{\mathbf{r}}\right)\label{eq:4.7}
\end{equation}
 and
\begin{eqnarray}
\int_{0}^{\infty}x^{\nu}j_{1}\left(x\right)dx & = & \lim_{\tau\rightarrow0^{+}}\int_{0}^{\infty}e^{-\tau x}x^{\nu}j_{1}\left(x\right)dx\nonumber \\
 & = & -\nu\cos\left(\frac{\pi}{2}\nu\right)\Gamma\left(\nu-1\right)\label{eq:4.8}
\end{eqnarray}
 for $\nu>-2$ , after a similar procedure as that of the $s$-wave
case, we obtain
\begin{multline}
\int d\mathbf{r}\frac{\mathcal{A}\left(\mathbf{X},{\bf r}_{j}+{\bf r}/2\right)Y_{1m}\left(\hat{\mathbf{r}}\right)}{r^{2}}e^{-i\mathbf{k}\cdot\mathbf{r}}\\
=-i\mathcal{A}\left(\mathbf{X},\mathbf{r}_{j}\right)Y_{1m}\left(\hat{\mathbf{k}}\right)\frac{4\pi}{k}+\alpha_{m}\left(\mathbf{X},\mathbf{r}_{j},\hat{\mathbf{k}}\right)\frac{2\pi}{k^{2}}\\
+i\beta_{m}\left({\bf X},{\bf r}_{j},\hat{{\bf k}}\right)\frac{\pi}{2k^{3}}+\mathcal{O}\left(k^{-4}\right),\label{eq:4.9}
\end{multline}
\begin{multline}
\int d\mathbf{r}\mathcal{B}\left(\mathbf{X},{\bf r}_{j}+{\bf r}/2\right)Y_{1m}\left(\hat{\mathbf{r}}\right)e^{-i\mathbf{k}\cdot\mathbf{r}}\\
=-i\mathcal{B}\left(\mathbf{X},\mathbf{r}_{j}\right)Y_{1m}\left(\hat{\mathbf{k}}\right)\frac{8\pi}{k^{3}}+\mathcal{O}\left(k^{-4}\right),\label{eq:4.10}
\end{multline}
and
\begin{equation}
\int d\mathbf{r}\mathcal{C}\left(\mathbf{X},{\bf r}_{j}+{\bf r}/2\right)rY_{1m}\left(\hat{\mathbf{r}}\right)e^{-i\mathbf{k}\cdot\mathbf{r}}=0,\label{eq:4.11}
\end{equation}
 where\begin{widetext}
\begin{eqnarray}
\alpha_{m}\left(\mathbf{X},{\bf R},\hat{\mathbf{k}}\right) & = & k^{2}\nabla_{{\bf R}}\mathcal{A}\left({\bf X},{\bf R}\right)\cdot\left[\nabla_{{\bf k}}\frac{Y_{1m}\left(\hat{{\bf k}}\right)}{k}\right],\label{eq:4.12}\\
\beta_{m}\left(\mathbf{X},{\bf R},\hat{\mathbf{k}}\right) & = & k^{3}\left(\nabla_{{\bf R}}\cdot\nabla_{{\bf k}}\right)\left[\nabla_{{\bf R}}\mathcal{A}\left({\bf X},{\bf R}\right)\cdot\nabla_{{\bf k}}\frac{Y_{1m}\left(\hat{{\bf k}}\right)}{k}\right],\label{eq:4.13}
\end{eqnarray}
and $\alpha_{m}$ and $\beta_{m}$ are independent on the amplitude
of ${\bf k}$ . The coupling term $\mathbf{r}\cdot\mathbf{L}\left({\bf X},{\bf r}_{j}+{\bf r}/2\right)$
contributes nothing to the tail of the large momentum distribution.
Inserting Eqs. (\ref{eq:4.9}), (\ref{eq:4.10}), and (\ref{eq:4.11})
into the expression of the momentum distribution Eqs (\ref{eq:2.26})
and (\ref{eq:2.27}), we obtain
\begin{multline}
n\left(\mathbf{k}\right)\sim32\pi^{2}\mathcal{I}_{a}\sum_{mm^{\prime}}G_{m}G_{m^{\prime}}^{*}Y_{1m}\left(\hat{\mathbf{k}}\right)Y_{1m^{\prime}}^{*}\left(\hat{\mathbf{k}}\right)\cdot\frac{1}{k^{2}}+\text{\text{{\bf Im}}}\left[\mathcal{N}\int d\mathbf{X}d\mathbf{R}\sum_{mm^{\prime}}G_{m}G_{m^{\prime}}^{*}\mathcal{A}Y_{1m}\left(\hat{\mathbf{k}}\right)\alpha_{m^{\prime}}^{*}\right]\cdot\frac{32\pi^{2}}{k^{3}}\\
+\left\{ 128\pi^{2}\mathcal{I}_{b}\sum_{mm^{\prime}}G_{m}G_{m^{\prime}}^{*}Y_{1m}\left(\hat{\mathbf{k}}\right)Y_{1m^{\prime}}^{*}\left(\hat{\mathbf{k}}\right)+8\pi^{2}\mathcal{N}\text{{\bf Re}}\sum_{mm^{\prime}}G_{m^{\prime}}^{*}G_{m}\int d\mathbf{X}d\mathbf{R}\left[\alpha_{m^{\prime}}^{*}\alpha_{m}-\mathcal{A}Y_{1m}\left(\hat{{\bf k}}\right)\beta_{m^{\prime}}^{*}\right]\right\} \cdot\frac{1}{k^{4}}+\mathcal{O}\left(k^{-5}\right),\label{eq:4.14}
\end{multline}
\end{widetext}where we have rewritten the integral variable $\mathbf{r}_{j}$
as $\mathbf{R}$. From the above equation, we can see that the $p$-wave
momentum distribution $n\left(\mathbf{k}\right)$ is not only dependent
on the amplitude of $\mathbf{k}$, but also on the direction of $\mathbf{k}$.
Therefore, the contact tensors, i.e., $\mathcal{C}_{a}^{(mm^{\prime})}$
and $\mathcal{C}_{b}^{(mm^{\prime})}$, may be introduced, if we use
the coefficients of the leading and subleading orders to define contacts,
just as what has been done in the work \cite{Yoshida2016P} for $\mathcal{C}_{a}^{(mm^{\prime})}$.
If we are only interested in the dependence of $n\left(\mathbf{k}\right)$
on the amplitude $k$, the expression can be simplified by integrating
$n\left(\mathbf{k}\right)$ over the direction of $\mathbf{k}$, and
we find that all the odd-order terms of $k^{-1}$ vanish. We finally
arrive at \begin{widetext}
\begin{equation}
n\left(k\right)\sim\frac{\sum_{m}\mathcal{C}_{a}^{(m)}}{k^{2}}+\left[\sum_{m}\mathcal{C}_{b}^{(m)}+8\pi^{2}\mathcal{N}\text{{\bf Re}}\sum_{mm^{\prime}}G_{m^{\prime}}^{*}G_{m}\int d\mathbf{X}d\mathbf{R}\int d\hat{{\bf k}}\left[\alpha_{m^{\prime}}^{*}\alpha_{m}-\mathcal{A}Y_{1m}\left(\hat{{\bf k}}\right)\beta_{m^{\prime}}^{*}\right]\right]\cdot\frac{1}{k^{4}}+\mathcal{O}\left(k^{-6}\right),\label{eq:4.16}
\end{equation}
\end{widetext}where the contacts $\mathcal{C}_{a}^{(m)}$ and $\mathcal{C}_{b}^{(m)}$
are defined as
\begin{eqnarray}
\mathcal{C}_{a}^{(m)} & \equiv & 32\pi^{2}\mathcal{I}_{a}\left|G_{m}\right|^{2},\label{eq:4.17}\\
\mathcal{C}_{b}^{(m)} & \equiv & 128\pi^{2}\mathcal{I}_{b}\left|G_{m}\right|^{2},\label{eq:4.18}
\end{eqnarray}
which are simply the diagonal elements of $\mathcal{C}_{a}^{(mm^{\prime})}$
and $\mathcal{C}_{b}^{(mm^{\prime})}$, if the contact tensors are
introduced. According to the definitions of $\mathcal{C}_{a}^{(m)}$
and $\mathcal{C}_{b}^{(m)}$, the adiabatic energy theorems (\ref{eq:3.8})
and (\ref{eq:3.9}) can alternatively be rewritten as
\begin{eqnarray}
\frac{\partial E}{\partial a_{m}^{-1}} & = & -\frac{\hbar^{2}\mathcal{C}_{a}^{(m)}}{32\pi^{2}M},\label{eq:4.19}\\
\frac{\partial E}{\partial b_{m}} & = & \frac{\hbar^{2}\mathcal{C}_{b}^{(m)}}{128\pi^{2}M},\label{eq:4.20}
\end{eqnarray}
which are consistent with the earlier definitions adopted in the previous
work \cite{Yu2015U,Yu2016U}, if one takes into account the difference
in the definitions of the effective range between ours and \cite{Yu2015U}. 

However, in the subleading behavior of the large momentum distribution,
there is an important difference. A new term appears, in addition
to the contact $\mathcal{C}_{b}^{(m)}$, similar to what occurs in
the $s$-wave case. This extra term is obviously resulted from the
c.m. motion of the pairs. Unfortunately, this crucial contribution
was omitted in the previous work \cite{Yu2015U}. At zero temperature,
it is perhaps a reasonable approximation to neglect this extra term,
as the c.m. motion of the pairs is likely frozen. However, at finite
temperature, it may contribute significantly, i.e., at the same order
as $\mathcal{C}_{b}^{(m)}$ in magnitude. As a result, the experimentally
observed subleading behavior of the large momentum distribution may
not be simply explained by the \emph{predicted} contact $\mathcal{C}_{b}^{(m)}$,
which is calculated theoretically by using the adiabatic energy theorem
Eq. (\ref{eq:4.20}).

Indeed, in the recent experiment by Luciuk \textit{et al.} \cite{Luciuk2016E},
both the leading and subleading behavior of the large momentum distribution
of a $p$-wave resonantly interacting Fermi gas of $^{40}$K atoms
were measured. A simple theoretical model, based on the adiabatic
energy theorems Eqs. (\ref{eq:4.19}) and (\ref{eq:4.20}), was used
to understand the experimental data. While the leading behavior is
reasonably explained using the calculated $\mathcal{C}_{a}^{(m)}$,
the subleading behavior can not be described using the calculated
$\mathcal{C}_{b}^{(m)}$ from the same model \cite{Luciuk2016E}.
The very existence of the extra term in the subleading behavior -
as we focused in this work - may give a possible reason for the disagreement. 

\section{Conclusions\label{sec:Conclusions}}

To conclude, we have systematically derived the adiabatic energy theorems
as well as the large momentum distribution of a polarized Fermi gas
with $p$-wave interactions, adopting the same approach used earlier
by Tan in his seminal works \cite{Tan2008E,Tan2008L,Tan2008G}. Two
$p$-wave contacts have been defined by connecting the tails of the
large momentum distribution to the adiabatic energy theorems, following
Tan's original idea of defining the contact for the $s$-wave interatomic
interaction. We have predicted that there is an extra term appearing
in the subleading behavior ($k^{-4}$) of the large momentum distribution
besides the contact related to the effective range, which has unfortunately
been omitted in the previous work \cite{Yu2015U}. This extra term
is associated with the center-of-mass motion of the interacting pairs,
and should be taken into account at finite temperature, and even at
zero temperature near the resonance with large pair fluctuations.

We believe that the existence of such an extra term - or alternatively
a new universal parameter - in the subleading behavior of the large
momentum distribution of strongly interacting Fermi gases is a general
feature, due to the introduction of the effective range of interactions,
which is necessary to regularize the higher-partial-wave interatomic
interactions. As a result, we cannot fully determine the short-distance,
large-momentum, or high-frequency behavior of correlation functions,
by simply defining some contacts through the adiabatic energy theorems,
as anticipated in the previous work \cite{Yu2015U}. At this point,
the full set of $p$-wave universal Tan relations remains to be amended.
On the other hand, the temperature dependence of the extra term or
the new universal parameter is to be understood.

\textit{Note added}. \textemdash{} We have recently become aware of
an erratum by Yu \emph{et al}. published in Physical Review Letters
\cite{Yu2016U}, in which the contribution from the center-of-mass
motion of the pairs to the subleading asymptotic behavior ($\sim k^{-4}$)
is included within their original theoretical frame.
\begin{acknowledgments}
We gratefully acknowledge valuable discussions with Shina Tan for
the normalization of the two-body wavefunction. SGP is supported by
the NKRDP (National Key Research and Development Program) under Grant
No. 2016YFA0301503 and NSFC under Grant No. 11474315. XJL and HH are
supported by the ARC Discovery Projects under Grant Nos. DP140100637,
FT140100003, FT130100815 and DP140103231. 

Correspondence should be addressed to SGP at pengshiguo@gmail.com.
\end{acknowledgments}

\begin{widetext}

\appendix

\section*{Appendix A: Derivation of $\int_{\mathcal{D}_{\epsilon}}\left|\Psi\left(\mathbf{X},\mathbf{R},\mathbf{r}\right)\right|^{2}d\mathbf{X}d\mathbf{R}d\mathbf{r}$}

In this appendix, we calculate $\int_{\mathcal{D}_{\epsilon}}\left|\Psi\left(\mathbf{X},\mathbf{R},\mathbf{r}\right)\right|^{2}d\mathbf{X}d\mathbf{R}d\mathbf{r}$,
where $\mathbf{r}$ and $\mathbf{R}$ are respectively the relative
and center-of-mass coordinates of the interacting pair of fermions
$\left(i,j\right)$, and ${\bf X}$ represents all the coordinates
of the rest of the fermions. Here, the domain $\mathcal{D}_{\epsilon}$
is all the configurations $\left(\mathbf{r}_{i},\mathbf{r}_{j}\right)$
in which $r=\left|\mathbf{r}_{i}-\mathbf{r}_{j}\right|>\epsilon$,
and $\epsilon$ is the finite range of the interaction potential between
fermions. Since we consider the situation that only the fermions $\left(i,j\right)$
interact with each other, while no more other pairs are close, the
many-body wavefunction can approximately be written as Eq. (\ref{eq:2.1}),
i.e., $\Psi\left(\mathbf{X},{\bf R},{\bf r}\right)\approx\mathcal{A}\left({\bf X},{\bf R}\right)\psi\left({\bf r}\right)$
with $\mathcal{A}\left({\bf X},{\bf R}\right)$ being a regular function.
Then
\begin{equation}
\int_{\mathcal{D}_{\epsilon}}\left|\Psi\left(\mathbf{X},\mathbf{R},\mathbf{r}\right)\right|^{2}d\mathbf{X}d\mathbf{R}d\mathbf{r}\approx\mathcal{I}_{a}\int_{r>\epsilon}d\mathbf{r}\left|\psi\left(\mathbf{r}\right)\right|^{2},\label{eq:A1}
\end{equation}
and $\mathcal{I}_{a}$ is already defined in Eq. (\ref{eq:2.14}).
Therefore, it turns out that we only need to consider a two-body problem,
and calculate the probability of finding the two particles outside
the interaction range $\epsilon$.

For a realistic two-body interaction potential with the finite range
$\epsilon$, the relative wavefunctions of two fermions corresponding
to different energies should be orthogonal, i.e.,
\begin{equation}
\int d\mathbf{r}\psi_{\mathbf{k}^{\prime}}^{*}\left(\mathbf{r}\right)\psi_{\mathbf{k}}\left(\mathbf{r}\right)=\delta_{\mathbf{k}^{\prime}\mathbf{k}},\label{eq:A2}
\end{equation}
where $\mathbf{k}^{\prime}$ and $\mathbf{k}$ are two relative wavevectors.
However, the wavefunction obtained from the pseudopotential method
is singular at the origin $\mathbf{r}=0$, since only the behavior
of the wavefunction outside the interaction potential, i.e., $r>\epsilon$,
is well-defined, and that inside the potential is not correctly described.
To see this, let us check the scalar product of $\psi_{\mathbf{k}^{\prime}}^{*}\left(\mathbf{r}\right)$
and $\psi_{\mathbf{k}}\left(\mathbf{r}\right)$ outside the range
of the interaction potential. We easily obtain from the Schr\"{o}dinger
equation that
\begin{equation}
\int_{r>\epsilon}d\mathbf{r}\psi_{\mathbf{k}^{\prime}}^{*}\left(\mathbf{r}\right)\psi_{\mathbf{k}}\left(\mathbf{r}\right)=\frac{\epsilon^{2}}{k^{2}-k^{\prime2}}\int_{r=\epsilon}d\hat{{\bf r}}\left[\psi_{\mathbf{k}^{\prime}}^{*}\left(\mathbf{r}\right)\frac{\partial\psi_{\mathbf{k}}\left(\mathbf{r}\right)}{\partial r}-\psi_{\mathbf{k}}\left(\mathbf{r}\right)\frac{\partial\psi_{\mathbf{k}^{\prime}}^{*}\left(\mathbf{r}\right)}{\partial r}\right],\label{eq:A3}
\end{equation}
where the integral on the right-hand side is over the surface $r=\epsilon$.
The $l$-th partial wavefunction obtained from the pseudopotential
takes the form
\begin{equation}
\psi_{\mathbf{k}}\left(\mathbf{r}\right)=\sum_{m}G_{m}k^{l+1}\left[\cot\delta_{lm}\cdot j_{l}\left(kr\right)-n_{l}\left(kr\right)\right]Y_{lm}\left(\hat{\mathbf{r}}\right),\label{eq:A4}
\end{equation}
which is in principle applicable even in the limit of $r=0$. Here,
$\delta_{lm}$ is the $l$-th partial wave scattering phase shift.
By inserting Eq. (\ref{eq:A4}) into Eq. (\ref{eq:A3}), we find that
the scalar product of $\psi_{\mathbf{k}^{\prime}}^{*}$ and $\psi_{\mathbf{k}}$
(\ref{eq:A3}) is divergent as $1/\epsilon^{2l-1}$ in the zero-range
limit $\epsilon\rightarrow0^{+}$ except the $s$-wave case ($l=0$).
This is not in agreement with the result obtained from a realistic
two-body potential, which should vanish as $\epsilon\rightarrow0$.
This unphysical result is due to the use of the pseudopential, and
then the two-body wavefunction obtained from the pseudopotential approach
should be normalized. For this purpose, we divide the integral $\int d\mathbf{r}\psi_{\mathbf{k}^{\prime}}^{*}\left(\mathbf{r}\right)\psi_{\mathbf{k}}\left(\mathbf{r}\right)$
into two parts as
\begin{equation}
\int d\mathbf{r}\psi_{\mathbf{k}^{\prime}}^{*}\left(\mathbf{r}\right)\psi_{\mathbf{k}}\left(\mathbf{r}\right)=\int_{r<\epsilon}d\mathbf{r}\psi_{\mathbf{k}^{\prime}}^{*}\left(\mathbf{r}\right)\psi_{\mathbf{k}}\left(\mathbf{r}\right)+\int_{r>\epsilon}d\mathbf{r}\psi_{\mathbf{k}^{\prime}}^{*}\left(\mathbf{r}\right)\psi_{\mathbf{k}}\left(\mathbf{r}\right)=0\label{eq:A5}
\end{equation}
for different $\mathbf{k}$ and $\mathbf{k}^{\prime}$, which in turn
gives
\begin{equation}
\int_{r<\epsilon}d\mathbf{r}\psi_{\mathbf{k}^{\prime}}^{*}\left(\mathbf{r}\right)\psi_{\mathbf{k}}\left(\mathbf{r}\right)=-\int_{r>\epsilon}d\mathbf{r}\psi_{\mathbf{k}^{\prime}}^{*}\left(\mathbf{r}\right)\psi_{\mathbf{k}}\left(\mathbf{r}\right).\label{eq:A6}
\end{equation}

\subsection{$s$-wave}

For the $s$-wave interaction with $l=0$, we find
\begin{eqnarray}
\int_{r<\epsilon}d\mathbf{r}\left|\psi_{\mathbf{k}}\left(\mathbf{r}\right)\right|^{2} & = & -\lim_{\mathbf{k}^{\prime}\rightarrow\mathbf{k}}\int_{r>\epsilon}d\mathbf{r}\psi_{\mathbf{k}^{\prime}}^{*}\left(\mathbf{r}\right)\psi_{\mathbf{k}}\left(\mathbf{r}\right)\nonumber \\
 & = & -4\pi\frac{\partial}{\partial k^{2}}\left(k\cot\delta_{0}\right)\nonumber \\
 & = & -2\pi b,\label{eq:A7}
\end{eqnarray}
where we have used Eq.(\ref{eq:A3}), and also set $\left|G_{0}\right|^{2}=4\pi$
without loss of generality. This means that the probability of two
atoms getting as close as $r<\epsilon$ should be $-2\pi b$, where
the effective range $b$ should be \emph{negative} in order to guarantee
the \emph{positive} probability inside the interaction potential.
Apparently, this is an alternative presentation of the well-known
Wigner's bound on the effective range, which was firstly derived by
Wigner from the causality and unitarity \cite{Wigner1955L}, and then
studied by Phillips and Cohen concerning the nucleon-nucleon scattering
\cite{Phillips1997H}. Subsequently, the probability of two atoms
outside the two-body potential should be
\begin{equation}
\int_{r>\epsilon}d\mathbf{r}\left|\psi_{\mathbf{k}}\left(\mathbf{r}\right)\right|^{2}=\int d\mathbf{r}\left|\psi_{\mathbf{k}}\left(\mathbf{r}\right)\right|^{2}+2\pi b.\label{eq:A8}
\end{equation}
This yields a well-normalized many-body wavefunction, 
\begin{equation}
\int_{\mathcal{D}_{\epsilon}}\left|\Psi\left(\mathbf{X},\mathbf{R},\mathbf{r}\right)\right|^{2}d\mathbf{X}d\mathbf{R}d\mathbf{r}\approx1+2\pi\mathcal{I}_{a}b.\label{eq:A9}
\end{equation}
In the zero-range limit, i.e., $\epsilon\rightarrow0^{+}$, the correction
from the $s$-wave effective range may reasonably be ignored due to
$b\sim0^{-}$.

\subsection{$p$-wave}

However, the situation is quite different for the higher partial waves.
For example, for the $p$-wave interaction with $l=1$, we know
\begin{equation}
\psi_{\mathbf{k}}\left(\mathbf{r}\right)=\sum_{m=-1}^{1}G_{m}k^{2}\left[\cot\delta_{1m}\cdot j_{1}\left(kr\right)-n_{1}\left(kr\right)\right]Y_{1m}\left(\hat{\mathbf{r}}\right),\label{eq:A10}
\end{equation}
and hence
\begin{eqnarray}
\int_{r<\epsilon}d\mathbf{r}\left|\psi_{\mathbf{k}}\left(\mathbf{r}\right)\right|^{2} & = & -\lim_{\mathbf{k}^{\prime}\rightarrow\mathbf{k}}\int_{r>\epsilon}d\mathbf{r}\psi_{\mathbf{k}^{\prime}}^{*}\left(\mathbf{r}\right)\psi_{\mathbf{k}}\left(\mathbf{r}\right)\nonumber \\
 & = & -\lim_{\mathbf{k}^{\prime}\rightarrow\mathbf{k}}\frac{\epsilon^{2}}{k^{2}-k^{\prime2}}\int_{r=\epsilon}d\hat{\mathbf{r}}\left[\psi_{\mathbf{k}^{\prime}}^{*}\left(\mathbf{r}\right)\frac{\partial\psi_{\mathbf{k}}\left(\mathbf{r}\right)}{\partial r}-\psi_{\mathbf{k}}\left(\mathbf{r}\right)\frac{\partial\psi_{\mathbf{k}^{\prime}}^{*}\left(\mathbf{r}\right)}{\partial r}\right]\nonumber \\
 & = & -\sum_{m}\left|G_{m}\right|^{2}\left[\frac{1}{\epsilon}+\frac{\partial}{\partial k^{2}}\left(k^{3}\cot\delta_{1m}\right)\right]\nonumber \\
 & = & -\sum_{m}\left|G_{m}\right|^{2}\left(\frac{b_{m}}{2}+\frac{1}{\epsilon}\right).\label{eq:A11}
\end{eqnarray}
It is clear that the $p$-wave effective range should satisfy $b_{m}\le-2/\epsilon$.
This is the simple generalization of Wigner's bound on the $p$-wave
effective range \cite{Hammer2009C,Hammer2010C,Braaten2012R}. We can
see that the $p$-wave effective range is driven to $-\infty$ if
we try to take the zero-range limit, i.e., $\epsilon\rightarrow0^{+}$,
in order to guarantee a positive probability inside the two-body potential
\cite{Pricoupenko2006P,Pricoupenko2006M,Nishida2012I}. Then the probability
of two atoms outside the two-body $p$-wave potential should be
\begin{equation}
\int_{r>\epsilon}d\mathbf{r}\left|\psi_{\mathbf{k}}\left(\mathbf{r}\right)\right|^{2}=\int d\mathbf{r}\left|\psi_{\mathbf{k}}\left(\mathbf{r}\right)\right|^{2}+\sum_{m}\left|G_{m}\right|^{2}\left(\frac{b_{m}}{2}+\frac{1}{\epsilon}\right),\label{eq:A12}
\end{equation}
and consequently,
\begin{equation}
\int_{\mathcal{D}_{\epsilon}}\left|\Psi\left(\mathbf{X},\mathbf{R},\mathbf{r}\right)\right|^{2}d\mathbf{X}d\mathbf{R}d\mathbf{r}\approx1+\mathcal{I}_{a}\sum_{m}\left|G_{m}\right|^{2}\left(\frac{b_{m}}{2}+\frac{1}{\epsilon}\right).\label{eq:A13}
\end{equation}
\end{widetext}

\end{document}